\documentclass[prb,aps,twocolumn,superscriptaddress,showpacs]{revtex4}
\usepackage{amssymb}
\usepackage{amsmath}
\usepackage[dvips]{graphicx}
\usepackage{dcolumn}
\usepackage{verbatim}
\usepackage{bm}
\usepackage{amsfonts}

\begin{document}
\title{Implementing a topological quantum model with cavity lattice}
\author{Ze-Liang Xiang}
\affiliation{Department of Physics and State Key Laboratory of Surface Physics, Fudan University, Shanghai 200433, China}
\affiliation{Key Laboratory of Micro and Nano Photonic Structures (Ministry of Education), Fudan University, Shanghai, 200433, China}
%\email{09110190024@fudan.edu.cn}

\author{Ting Yu}
\affiliation{Department of Physics and Engineering Physics, Stevens Institute of Technology, Hoboken, New Jersey 07030-5991, USA}

\author{Wenxian Zhang}
\affiliation{Key Laboratory of Micro and Nano Photonic Structures (Ministry of Education), Fudan University, Shanghai, 200433, China}
\affiliation{Department of Optical Science and Engineering, Fudan University, Shanghai 200433, China}

\author{Xuedong Hu}
\affiliation{Department of Physics, University at Buffalo-SUNY, Buffalo, New York 14260-1500, USA}

\author{J. Q. You}
\affiliation{Department of Physics and State Key Laboratory of Surface Physics, Fudan University, Shanghai 200433, China}
\affiliation{Key Laboratory of Micro and Nano Photonic Structures (Ministry of Education), Fudan University, Shanghai, 200433, China}
\email{jqyou@fudan.edu.cn}

\date{\today}

\begin{abstract}
Kitaev model has both Abelian and non-Abelian anyonic
excitations. It can act as a starting point for topological quantum computation.
However, this model Hamiltonian is difficult to implement in natural
condensed matter systems. Here we propose a quantum simulation scheme
by constructing the Kitaev model Hamiltonian in a lattice of coupled cavities
with embedded $\Lambda$-type three-level atoms. In this scheme, several
parameters are tunable, for example, via external laser fields. Importantly, our
scheme is based on currently existing technologies and it provides a feasible
way of realizing the Kitaev model to explore topological excitations.
\end{abstract}
\maketitle

\section{Introduction}

Topological quantum systems are attracting considerable interest because of
their fundamental importance in diverse areas of physics, ranging from
quantum-field theory to semiconductor physics~\cite{XLQ,MZH}, as well as
emerging fields such as quantum computation~\cite{CN,AK1}. A topological
quantum system has topological phases of matter that are insensitive to local
perturbations. For example, a recently proposed quantum model with
quasi-local spin interactions possesses emergent topologically ordered
states~\cite{MAL}. It has also been suggested that a two-dimensional (2D)
electron gas in the fractional quantum Hall regime has non-Abelian anyonic
excitations near particular magnetic fields, which can be used for topological
quantum computing~\cite{TE,SDS}.

\begin{center}
\begin{figure}
\includegraphics[width=2.8in]{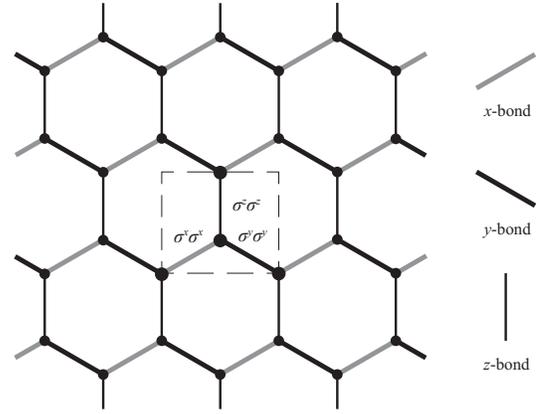}
\caption{Schematic diagram of the Kitaev model on a honeycomb lattice.  The building 
block in our scheme (denoted as a dashed box) is explicitly shown in Fig.~\ref{fig2}(a).} 
\label{fig1}
\end{figure}
\end{center}

In addition to the search for naturally existing topological phases, there
have also been wide-ranging theoretical efforts to design model Hamiltonians
and artificial lattice structures that possess nontrivial topological
properties. One of the prominent candidates for topological quantum 
computation is the Kitaev model on a honeycomb lattice~\cite{AK2}, as shown 
in Fig.~\ref{fig1}. This model can be described by a completely anisotropic 
Hamiltonian
\begin{equation}
H = J_x\sum_{x-{\rm link}}\sigma^x_i\sigma^x_j +J_y\sum_{y-{\rm link}}
\sigma^y_i\sigma^y_j + J_z\sum_{z-{\rm link}}\sigma^z_i\sigma^z_j \,,
\end{equation}
where $J_{x}$, $J_{y}$, and $J_{z}$ denote the coupling strengths of the
$x$-, $y$-, and $z$-type bonds, respectively. Using Majorana fermion
method~\cite{AK2} or Jordan-Wigner transformation~\cite{XYF,GK}, the Kitaev model can be
analytically solved. It has two phases: a band insulator phase and a topological
gapless phase~\cite{AK2}. In the insulator phase, which is equivalent to a toric code
model for the same braiding statistics~\cite{JRW}, the elementary excitations are
Abelian anyons. However, the vortices in the gapless phase are not
well-defined because the rule of the braiding statistics is unclear in this case.
An applied external magnetic field breaks the time reversal symmetry in the
Hamiltonian and opens a gap in the gapless phase.  Now the vortices obey a
well-defined non-Abelian anyonic statistics~\cite{AK2}. With possible non-Abelian
anyonic excitations, the Kitaev model could play an important role in
demonstrating anyonic statistics and in implementing topological quantum computing.

Due to its strong anisotropicity, the Kitaev model is unlikely to exist in a
natural homogeneous physical system. An alternative but promising way to realize
this model is to use artificial structures~\cite{IB}, such as Josephson junction
arrays~\cite{JQY}, optical lattices~\cite{LMD,CZ}, and polar
molecules~\cite{AM,GJ}. However, with a Josephson junction array, the magnetic
term of the model Hamiltonian cannot be cancelled completely~\cite{JQY}, while
with other proposals (in either optical lattice or molecules) extremely strict
conditions are required, such as ultra-low temperatures.

Recently, various theoretical proposals of using coupled cavities to simulate
basic spin-interactions and many-body models have been studied~\cite{DGA,MJH1,JC1,CZX,JC2,MJH2,ADG}. In
particular, the XY spin model and the Heisenberg model have been theoretically
simulated in reference~\cite{DGA} and reference~\cite{MJH1}, respectively. In reference~\cite{MJH1},
the $\sigma^x\sigma^x$ and $\sigma^y\sigma^y$ couplings were proposed to
implement by using two laser fields (here denoted as A and B), with the related
coupling strengths tuned by varying the applied laser fields. In addition, the
$\sigma^z \sigma^z$ coupling was proposed to implement by using another two laser
fields (here denoted as C and D). However, these three types of spin couplings
cannot occur simultaneously. To solve this problem, the Suzuki-Trotter formula
was employed in reference~\cite{MJH1}, which involves applying an appropriate repeated
sequence of pulses that tunes on the laser fields A and B for a short interval
of time after another interval of time in which the laser fields C and D are tuned
on. This can in principle yield an anisotropic Heisenberg Hamiltonian with these
three types of spin couplings.

In this paper, we propose an alternative approach to simultaneously implement all
three different couplings of spins in a quantum model. However, in sharp contrast
to the proposal in reference~\cite{MJH1}, the simultaneous implementation of these three
different spin couplings in our considered model is simply owing to the introduction
of suitable cavity modes depending on the geometrical structure of the cavity
lattice. Moreover, our proposal is also experimentally accessible with currently
existing technologies~\cite{BSS,TA,KH}. Specifically, we show that a lattice of
tunnel-coupled cavities can be tuned to emulate the strongly anisotropic Kitaev
lattice. In our case, the basic element is a $\Lambda$-type three-level atom inside
three cavities that are oriented in different directions, where the two long-lived
atomic levels form an effective spin-$\frac{1}{2}$ (qubit). The interactions between
nearest-neighbor atoms are realized via the exchange of virtual photons between the
coupled cavities. Indeed, our scheme for building qubits and achieving inter-qubit
interactions is quite general, so that complex many-body models can be constructed
by changing the structure of the cavity lattice and/or by varying the driving laser
fields, without making use of the Suzuki-Trotter formula that involves applying an
appropriate repeated sequence of pulses~\cite{MJH1}.

The paper is organized as follows. In Sec.~2, we describe the design of our cavity
lattice to simulate the Kitaev model. We derive the effective Hamiltonian of the
system in two different cases in Sec.~3 and then discuss the conditions for
experimental implementation in Sec.~4. Finally, we give a summary of our scheme
and our conclusions in Sec.~5.

\begin{center}
\begin{figure}[tbp]
\includegraphics[width=3.2in]{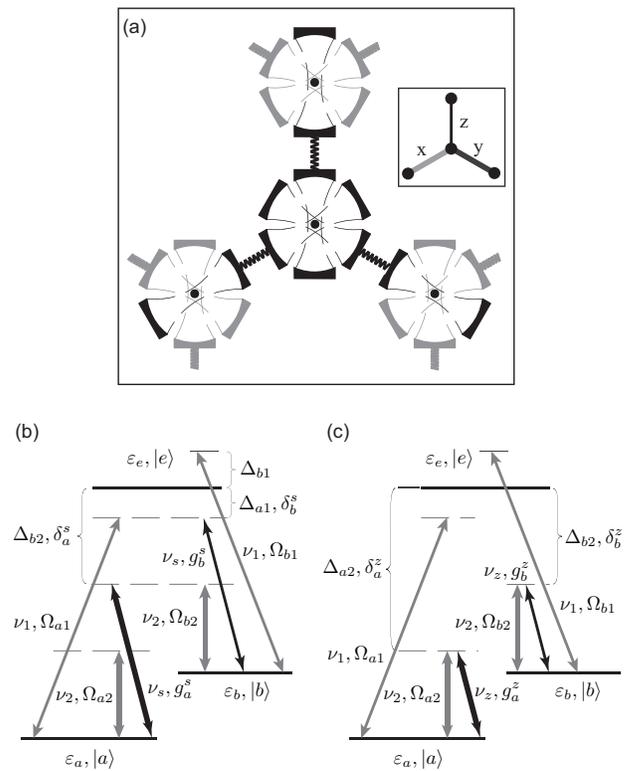}
\caption{(a) Schematic diagram of the building block for constructing the Kitaev lattice on a 
honeycomb lattice. At each site of the honeycomb lattice, there are three cavities 
oriented 120$^{\circ}$ apart, where a $\Lambda$-type three-level atom is placed in 
the common region of the three cavity modes belonging to different cavities.
(b) Schematic diagram of a $\Lambda$-type three-level atom coupled with two laser
fields (denoted as gray arrows) and a mode of the cavity (denoted as black arrows)
in the $s$-bond direction of the honeycomb lattice, where $s=x$ and $y$.
(c) Schematic diagram of a $\Lambda$-type three-level atom coupled with two laser
fields and a mode of the cavity in the $z$-bond direction of the honeycomb lattice.
In both (b) and (c), a laser field of frequency $\nu_1$ drives the transitions
$|a\rangle\leftrightarrow|e\rangle$ and $|b\rangle\leftrightarrow|e\rangle$ with
Rabi frequencies $\Omega_{a1}$ and $\Omega_{b1}$, respectively. The detuning
between $\nu_1$ and $\omega_{ea}\equiv\varepsilon_e-\varepsilon_a$
($\omega_{eb}\equiv\varepsilon_e-\varepsilon_b$) is $\Delta_{a1}$ ($\Delta_{b1}$).
Another laser field with frequency $\nu_2$ also drives the transitions
$|a\rangle\leftrightarrow|e\rangle$ and $|b\rangle\leftrightarrow|e\rangle$ with
Rabi frequencies $\Omega_{a2}$ and $\Omega_{b2}$, respectively. The detuning
between $\nu_2$ and $\omega_{ea}$ ($\omega_{eb}$) is $\Delta_{a2}$ ($\Delta_{b2}$).
Furthermore, a cavity mode of frequency $\nu_k$ is coupled to the transitions
$|a\rangle\leftrightarrow|e\rangle$ and $|b\rangle\leftrightarrow|e\rangle$ with
coupling strengths $g_a^k$ and $g_b^k$, respectively, where $k=x$, $y$ and $z$.
The detuning between $\nu_k$ and $\omega_{ea}$ ($\omega_{eb}$) is $\delta_a^k$
($\delta_b^k$).} 
\label{fig2}
\end{figure}
\end{center}

\section{Model}

We construct an artificial 2D honeycomb lattice using tunnel-coupled cavities to
implement the Kitaev model (see Figs.~\ref{fig1} and \ref{fig2}). The building block of this
lattice involves four lattice sites, as shown in the dashed box in Fig.~\ref{fig1}. At
each site, we place an identical $\Lambda$-type three-level atom, which is coupled
to three cavities [see Fig.~\ref{fig2}(a)]. The three-level atom has two long-lived ground
states, $|a\rangle$ and $|b\rangle$, which are denoted as spin-down and spin-up
states, respectively.  These effective spin-$\frac{1}{2}$'s form qubits.  The
$\sigma^x\sigma^x$, $\sigma^y\sigma^y$, and $\sigma^z \sigma^z$ couplings are
achieved by exchanging virtual photons between nearest-neighbor cavities in the
three bond directions of the Kitaev model.

The 2D lattice of cavities with embedded atoms is described by a Hamiltonian
consisting of three parts: $H_A$, $H_C$, and $H_{AC}$. Here $H_A$ is the
Hamiltonian of the bare atoms:
\begin{eqnarray}
H_A\; = \;\sum^N_{j=1} \left( \omega_{ea} \sigma^j_{ee} + \omega_{ba} \sigma^j_{bb} \right) \,,
\label{HA}
\end{eqnarray}
where $\sigma_{pq} = |p\rangle \langle q|$, with $p, q \in \{a,b,e \}$, $j$
is the lattice index, and $\omega_{ea}$ ($\omega_{ba}$) is the energy
difference between states $|e\rangle$ ($|b\rangle$) and $|a\rangle$. In
Eq.~(\ref{HA}), the energy of state $|a\rangle$ is chosen to be the zero point,
and we set $\hbar=1$.

$H_C$ is the Hamiltonian of the cavity lattice in the absence of atoms. In
contrast to early works~\cite{MJH1,JC1,CZX}, our model involves more cavity modes, and
thus $H_C$ can be described as
\begin{subequations}
\begin{eqnarray}
\!&\! \!&\!H_C = \sum_j^N\sum_{k=x,y,z}H_{C0}^{(k)}(j)+\sum_{k=x,y,z}H_{CJ}^{(k)}\,,\label{HC}\\
\!&\! \!&\!H_{C0}^{(k)}(j) =  \nu_k a^{\dag}_{kj} a_{kj}\,, \label{HC0} \\
\!&\!\!&\!H_{CJ}^{(k)}=\sum_{\langle i,j \rangle} t_k(a^{\dag}_{ki} a_{kj} + a_{ki}a^{\dag}_{kj})\,,
\label{HCJ}
\end{eqnarray}
\end{subequations}
where $a_{kj}$, with $k=x$, $y$ or $z$, is the annihilation operator of
photons, with frequency $\nu_k$, for a mode of the cavity at the $j$th
site, which is oriented in the $k$-bond direction. For simplicity, we
assume that all cavities in the same bond direction have the identical mode
frequency. $H_{C0}^{(k)}(j)$ is the Hamiltonian for cavity photons at the 
$j$th site of the lattice, and $H_{CJ}^{(k)}$ describes the tunnelling between
nearest-neighbor cavities along the $k$ direction, with tunnelling strength
$t_k$. Here the small interactions among photons of different cavity modes
are neglected, because the common region of the three cavity modes at each
site can be designed very small.

Lastly, $H_{AC}$ describes the interaction between atoms and photons:
\begin{subequations}
\begin{eqnarray}
H_{AC} \!&\! = \!&\! \sum_j^N\left[H_{{\rm int}}^{(1)}(j) + H_{{\rm int}}^{(2)}(j)\right]\,, \label{HAC}\\
H_{{\rm int}}^{(1)}(j) \!&\! = \!&\! \sum_{k=x,y,z}\sum_{l=a,b}\left(g^k_la_{kj}\sigma^j_{el}+{\rm H.c.}\right)\,, \label{HAC1} \\
H_{{\rm int}}^{(2)}(j) \!&\! = \!&\! \sum_{l=a,b}\left[\left(\frac{1}{2}\Omega_{l1}e^{-i\nu_1t}+\frac
{1}{2}\Omega_{l2}e^{-i\nu_2t} \right) \sigma^j_{el} + {\rm H.c.}\right]\,,\nonumber\\
\label{HAC2}
\end{eqnarray}
\end{subequations}
where $H_{{\rm int}}^{(i)}$, with $i=1~(2)$, is the interaction Hamiltonian
between the atoms and the cavity photons (the two laser fields). Each cavity
mode is coupled with transitions $|a\rangle\leftrightarrow|e\rangle$ and
$|b\rangle\leftrightarrow|e\rangle$, and these transitions are also driven by
two laser fields. In $H_{{\rm int}}^{(1)}$, $g^k_{a(b)}$ is the coupling
strength between the cavity mode $k$ and the atomic transition
$|a\rangle\leftrightarrow|e\rangle$ ($|b\rangle\leftrightarrow|e\rangle$). In
$H_{{\rm int}}^{(2)}$, $\Omega_{a1}$ ($\Omega_{a2}$) and $\Omega_{b1}$
($\Omega_{b2}$) are the Rabi frequencies involving the driving processes of
the transitions $|a\rangle\leftrightarrow|e\rangle$ and
$|b\rangle\leftrightarrow|e\rangle$ by the laser field of frequency $\nu_1$
($\nu_2$). Here we assume that $\omega_{ba}\gg g_a^k$, $g_b^k$, $\Omega_{a1}$,
$\Omega_{a2}$, $\Omega_{b1}$, and $\Omega_{b2}$.

In the following section, we show how to obtain the $\sigma^x\sigma^x$,
$\sigma^y\sigma^y$, and $\sigma^z\sigma^z$ couplings in the $x$, $y$, and $z$
directions simultaneously, so as to implement the Kitaev model using the cavity
lattice.

\section{Effective Hamiltonian}

\subsection{Reduced pseudo-spins}

For the atom at site $j$ of the honeycomb lattice, when the interaction picture
with respect to $H_0(j) =H_A(j) + \sum_{k=x,y,z} H_{C0}^{(k)}(j)$ is used,
$H_{{\rm int}}^{(1)}(j)$ and $H_{{\rm int}}^{(2)}(j)$ are transformed to
\begin{subequations}
\begin{eqnarray}
H_I^{(1)}(j) \!&\!=\!&\!\sum_{k=x,y,z}\sum_{l=a,b}\left(g^k_la_{kj}e^{i\delta^k_lt}\sigma^j_{el} + {\rm H.c.}\right)\,,\label{HI1}\\
H_I^{(2)}(j) \!&\!=\!&\! \sum_{l=a,b}\left[\left(\frac{1}{2}\Omega_{l1}e^{i\Delta_{l1}t}+\frac{1}{2}\Omega_{l2}e^{i\Delta_{l2}t}\right)\sigma^j_{el}+ {\rm H.c.}\right]\,,\nonumber\\
\label{HI2}
\end{eqnarray}
\end{subequations}
where the detunings between the laser frequencies and the energy differences of
the atomic transitions are given by $\Delta_{a1} = \omega_{ea} - \nu_1$,
$\Delta_{a2} = \omega_{ea} - \nu_2$, $\Delta_{b1} = \omega_{eb} - \nu_1$, and
$\Delta_{b2} = \omega_{eb} - \nu_2$. The detunings between the cavity-mode
frequencies and the energy differences of the atomic transitions are
$\delta^k_a = \omega_{ea} - \nu_k$ and $\delta^k_b = \omega_{eb} - \nu_k$. For
a more detailed description, see Figures.~2(b) and 2(c). In our scheme, all the 
detunings are red-shifted except for $\Delta_{b1}$. Here we assume that $\Delta_{a1}$,
$\Delta_{a2}$, $\Delta_{b1}$, $\Delta_{b2}$, $\delta^k_a$ and $\delta^k_b$ are
all in the large-detuning regime, i.e.,
$|\delta_{\alpha}^k|,|\Delta_{\beta}|\gg|g_{\alpha}^k|,\Omega_{\beta}$, where
$\alpha=a$ and $b$, and $\beta=a1$, $a2$, $b1$, and $b2$. In this regime, two
kinds of transitions can dominate when $\Delta_{a1}\sim\delta_b^s$,
$\Delta_{a2}\sim\delta_a^z$, $\Delta_{b2}\sim\delta_a^s$, and
$\Delta_{b2}\sim\delta_b^z$, where $s=x$ and  $y$. One is the Raman process, in
which the atom is excited from state $|a\rangle$ ($|b\rangle$) by absorbing a
single laser photon and then falls to state $|b\rangle$ ($|a\rangle$) by emitting
a single cavity photon and vice versa. The other is the Rayleigh process, in
which the atom is excited from state $|a\rangle$ ($|b\rangle$) by absorbing a
single laser photon and then returns to $|a\rangle$ ($|b\rangle$) by emitting a
single cavity photon and vice versa. Also, the Rayleigh process can occur by
absorbing a single laser (cavity) photon and then emitting a single laser (cavity)
photon.

Because the detunings shown above are all in the large-detuning regime, the
excited state can be adiabatically eliminated from the Hamiltonian as long as
the initial state of the atom is within the subspace spanned by the two
long-lived states $|a\rangle$ and $|b\rangle$. This yields the two long-lived
states to be shifted in energy and also be effectively coupled by the two-photon
Raman and Rayleigh processes. We first study the simple case [ see Figs.~\ref{fig2}(b)
and \ref{fig2}(c)]:
\begin{equation}
\Delta_{a1} = \delta_b^s,~~ \Delta_{a2} = \delta_a^z, ~~\Delta_{b2} = \delta_a^s = \delta_b^z.
\label{DEL0}
\end{equation}
With the dominant transitions considered, $H_I^{(1)}(j)+H_I^{(2)}(j)$ is reduced,
in the large detuning regime, to~\cite{DFVJ}
\begin{eqnarray}
H_e (j) \!&\! =\!&\!- \left( \eta_{a1} + \eta_{a2} \right) \sigma_{aa}^{j} - (\eta_{b1} + \eta_{b2}) \sigma_{bb}^{j} \nonumber\\
\!&\!\!&\! - \sum_{k=x,y,z}  \left(\lambda_a^k a^{\dag}_{kj} a_{kj} \sigma^j_{aa} + \lambda_b^k a^{\dag}_{kj} a_{kj} \sigma^j_{bb} \right) \nonumber\\
\!&\!\!&\! - \sum_{s=x,y} \left[ \left( A_{s1} \sigma^j_{ba} + A_{s2} \sigma^j_{ab} \right) a^{\dag}_{sj} + {\rm H.c.} \right] \nonumber\\
\!&\!\!&\! - \left[ \left( A_{z1} \sigma^j_{aa} + A_{z2} \sigma^j_{bb} \right) a^{\dag}_{zj} + {\rm H.c.} \right] \nonumber\\
\!&\!\!&\! +\, F(a_m^{\dag} a_n)\,, \label{HE1}
\end{eqnarray}
where $\eta_{\beta}\equiv\Omega_{\beta}^{2}/4\Delta_{\beta}$, with $\beta=a1$,
$a2$, $b1$, and $b2$; $\lambda_a^k \equiv (g_a^k)^2/\delta_a^k$, and
$\lambda_b^k \equiv (g_b^k)^2/\delta_b^k$;
\begin{eqnarray}
A_{s1} = \frac{g^s_b \Omega_{a1}}{2\delta_b^s}\,, \!&\!\quad\!&\!
A_{s2} = \frac{g^s_a\Omega_{b2}}{2\delta_a^s}\,,\nonumber\\
A_{z1} = \frac{g^z_a \Omega_{a2}}{2\delta_a^z}\,, \!&\!\quad\!&\!
A_{z2} = \frac{g^s_b\Omega_{b2}}{2\delta_b^z}\,.\nonumber
\end{eqnarray}
Here we assume that $\lambda_k\equiv\lambda_a^k=\lambda_b^k$. $F(a_m^{\dag}a_n)$
in the last line of Eq.~(\ref{HE1}) represents all the terms containing the operators
$a_m^{\dag}a_n$ ($m,n\in\{x,y,z\}$, and $m\neq n$), and describes the effective
interaction between different cavity modes via atoms. Note that the term
$F(a_m^{\dag}a_n)$ does not appear in the other proposals of  cavity-based
quantum simulations of spin models~\cite{DGA,MJH1,JC1,CZX}, because only one cavity
mode was used therein.

If $2|g_{\alpha}| \gg|\Omega_{\beta}|$, then
$\lambda_k \gg \Omega_{\beta}^2/4\Delta_{\beta}$. Thus, the AC Stark shift of two
long-lived states induced by the laser fields can be neglected. As a matter of
fact, when the condition $2|g_{\alpha}| \gg|\Omega_{\beta}|$ is not satisfied,
one can alternatively introduce other laser fields to compensate the AC Stark
effect~\cite{JC2}. In the large-detuning regime that we consider above, the zero-photon
subspace is preserved because only virtual two-photon processes are involved. As
shown in the next section, all the photonic degrees of freedom are eliminated in
the zero-photon subspace when implementing the second adiabatic elimination. Thus,
we can neglect the effective interaction between different cavity modes in Eq.~(\ref{HE1}),
because after the second adiabatic elimination, the terms arising from
$F(a_m^{\dag}a_n)$ can be guaranteed to vanish in the zero-photon subspace. To
briefly summarize, in this subsection we have shown that the three-level atom at
each site of the lattice can be reduced to an effective two-level system in the
large-detuning regime.

\subsection{Effective couplings between pseudo-spins}

\begin{center}
\begin{figure}
\includegraphics[width=2.8in]{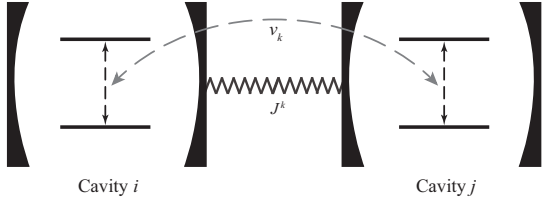}
\caption{Schematic diagram of the exchange of virtual photons between adjoining cavities
to induce an effective coupling between atoms at nearest-neighbor site of the
honeycomb lattice. The gray dashed curve with arrows denotes the virtual-photon
exchange.}
\label{fig3}
\end{figure}
\end{center}

Below we attempt to eliminate the photonic degrees of freedom and derive a pure
spin Hamiltonian for the pseudo-spins defined by the atomic states $|a\rangle$
and $|b\rangle$ of the original three-level atoms. We consider the interaction
picture with respect to $H^{\prime}_0=-\sum_{k=x,y,z}\lambda_k(a^{\dag}_{kj}
a_{kj}\sigma^j_{aa} - a^{\dag}_{kj} a_{kj} \sigma^j_{bb})$. When the tunnelling
term between nearest-neighbor cavities, $\sum_{k=x,y,z}\sum_{\langle i,j \rangle} t_k (a^{\dag}_{ki}a_{kj}+a_{ki}
a^{\dag}_{kj})$, is included, a new form of the Hamiltonian is obtained in this
interaction picture. Again, using the assumption of the large detuning
\begin{equation}
\lambda_k \gg \frac{|g_{\alpha}\Omega_{\beta}|}{2\Delta_{\beta}},~t_k\,, \label{2aec}
\end{equation}
we make the second adiabatic elimination~\cite{JC1}. Here the dominant process involves
only the exchange of virtual photons between nearest-neighbor cavities (as shown
in Fig.~\ref{fig3}), because other transitions, which oscillate rapidly with large
frequencies, can be neglected in the long-time limit. Finally, the effective spin
Hamiltonian in the zero-photon subspace can be written as
\begin{eqnarray}
H_{{\rm eff}}\!&\! =\!&\! \sum^N_j[B_x+B_y+B_z+\frac{1}{4}(-J_{z1}+J_{z2})]\sigma^z_j\nonumber\\
\!&\!\!&\!+ \sum_{x-{\rm link}}\frac{1}{2}[(J_{x1} + J_{x2}) \sigma^x_i\sigma^x_j+(J_{x1}-J_{x2})\sigma^y_i\sigma^y_j]\nonumber\\
\!&\!\!&\! + \sum_{y-{\rm link}}\frac{1}{2}[(J_{y1}+J_{y2})\sigma^x_i\sigma^x_j+(J_{y1}-J_{y2})\sigma^y_i\sigma^y_j] \nonumber\\
\!&\!\!&\! +\sum_{z-{\rm link}}\frac{1}{4}(J_{z1}+J_{z2}-2J_{z3})\sigma^z_i\sigma^z_j\,, \label{HE2}
\end{eqnarray}
with
\begin{eqnarray}
B_x\!&\!=\!&\!B_y=\frac{1}{2}(\eta_{b2}-\eta_{a1})\,,~B_z=\frac{1}{2}(\eta_{b2}-\eta_{a2})\,,\nonumber\\
J_{s1}\!&\!=\!&\!\frac{t_s}{4}\left[\left(\frac{\Omega_{a1}}{g_b^s}\right)^2+\left(\frac{\Omega_{b2}}{g_a^s}\right)^2\right]\,,\nonumber\\
J_{s2}\!&\!=\!&\!\frac{t_s}{2}\left(\frac{\Omega_{a1}\Omega_{b2}}{g_a^sg_b^s}\right)\,,~~J_{z1}=\frac{t_z}{2}\left(\frac{\Omega_{a2}}{g_a^z}\right)^2\,,\nonumber\\
J_{z2}\!&\!=\!&\!\frac{t_z}{2}\left(\frac{\Omega_{b2}}{g_b^z}\right)^2\,,~~J_{z3} = \frac{t_z}{2}\left(\frac{\Omega_{a2}\Omega_{b2}}{g_a^zg_b^z}\right)\,. \nonumber
\end{eqnarray}
where $s=x$, and $y$.

We define $\gamma_k\equiv\delta_a^k/\delta_b^k$, so $\gamma_x=\gamma_y$ in this
simple case. If the parameters are properly chosen as listed in Table.~1, the
energy shifts of the two long-lived states, which generate from the adiabatic
elimination, can cancel with each other, and the coefficient of the magnetic
term thereby becomes zero. Furthermore, these conditions lead to $J_{x1}=J_{x2}$,
$J_{y1}=-J_{y2}$, and $J_{z1}=J_{z2}=-J_{z3}$, so the effective Hamiltonian is
reduced to the Kitaev model on a honeycomb lattice:
\begin{equation}
H_{{\rm eff}}=J_x\sum_{x-{\rm link}}\sigma^x_i\sigma^x_j + J_y\sum_{y-{\rm link}}\sigma^y_i\sigma^y_j + J_z\sum_{z-{\rm link}}\sigma^z_i\sigma^z_j\,, \label{HEF}
\end{equation}
where
\begin{equation}
J_x = \frac{t_x}{2} \left(\frac{\Omega_{b2}}{g_a^x}\right)^2\,,~~
J_y = \frac{t_y}{2} \left(\frac{\Omega_{b2}}{g_a^y}\right)^2\,,~~
J_z = \frac{t_z}{2} \left(\frac{\Omega_{b2}}{g_b^z}\right)^2\,\nonumber
\end{equation}
are the $x$-,$y$- and $z$-type coupling strengths between atoms at
nearest-neighbor sites when Eq.~(\ref{DEL0}) is satisfied. So far, the three different
spin couplings in the Kitaev model have been emulated simultaneously by
introducing three different cavity modes depending on the geometrical structure
of the honeycomb lattice.

Furthermore, by properly changing the Rabi frequency $\Omega_{a2}$ in the last
column of Table.~1, a new magnetic term can be generated:
\begin{eqnarray}
H_{{\rm eff}} \!&\!=\!&\! B\sum_j\sigma^z_j + J_x\sum_{x-{\rm link}}\sigma_i^x\sigma_j^x\nonumber\\
\!&\!\!&\!+J_y\sum_{y-{\rm link}} \sigma_i^y\sigma_j^y+J_{zc}\sum_{z-{\rm link}}\sigma_i^z\sigma_j^z\,, \label{HEFB}
\end{eqnarray}
where $B=B_{z}-(J_{z1}-J_{z2})/4$, and $J_{zc}=(J_{z1}+J_{z2}-2J_{z3})/4$.
With this effective Hamiltonian, it is possible to investigate the phase
transition from a gapless phase of the Kitaev model to a gapped phase when
the effective magnetic field is presented.

\begin{center}
\begin{tabular}{p{17.5mm}p{27.5mm}p{27.5mm}}
\multicolumn{3}{p{80mm}}{{\bf Table~1}\quad Conditions of the
parameters for implementing the Kitaev model in the case when Eq.~(\ref{DEL0})
is satisfied.}\\
\hline\hline
Parameters & \quad$x$-type bond $~~~$ $~~~~~~~$($\sigma^x\sigma^x$) & \quad$y$-type bond $~~~$ $~~~~~~~$($\sigma^y\sigma^y$) \\
\hline
\quad$\Omega_{a1}$ & \multicolumn{2}{c}{} \\
\quad$\Omega_{b2}$ & \multicolumn{2}{c}{\raisebox{1.8ex}[0pt]{$\gamma_s\Omega^2_{a1}=\Omega^2_{b2}$}\quad\quad} \\
\quad$\Omega_{a2}$ &\quad\quad\quad ------ &\quad\quad\quad ------ \\
\quad$g^k_a$ & \quad\quad$g_a^xg_b^x>0$ & \quad\quad\quad$g_a^yg_b^y<0$ \\
\quad$g^k_b$ & \multicolumn{2}{c}{$(g^k_a)^2=\gamma_k(g^k_b)^2$\quad\quad} \\
\hline\hline
\\
\hline\hline
Parameters & \multicolumn{2}{c}{$z$-type bond ($\sigma^z\sigma^z$)\quad\quad} \\
\hline
\quad$\Omega_{a1}$ & \multicolumn{2}{c}{------\quad\quad\quad} \\
\quad$\Omega_{b2}$ & \multicolumn{2}{c}{} \\
\quad$\Omega_{a2}$ & \multicolumn{2}{c}{\raisebox{1.8ex}[0pt]{$\Omega^2_{a2}=\gamma_z\Omega^2_{b2}$}\quad\quad\quad} \\
\quad$g^k_a$ & \multicolumn{2}{c}{$g_a^zg_b^z<0$\quad\quad\quad} \\
\quad$g^k_b$ & \multicolumn{2}{c}{$(g^k_a)^2=\gamma_k(g^k_b)^2$\quad\quad\quad} \\
\hline\hline
\end{tabular}
\end{center}

\subsection{More generic effective Hamiltonians}

In the above subsections, we have derived the effective Hamiltonians in the case
shown in Figs.~\ref{fig2}(b) and \ref{fig2}(c). In this subsection, we consider a more general
case where the detunings for the driving laser fields are not equal to the
related detunings for the cavity modes, i.e., $\Delta_{a1} \neq \delta_b^s$,
$\Delta_{a2} \neq \delta_a^z$, and $\Delta_{b2} \neq \delta_a^s = \delta_b^z$,
as illustrated in Figure~4. The following new notations
\begin{equation}
\delta_{x(y)}\equiv\delta^{x(y)}_a - \Delta_{b2} = \delta^{x(y)}_b - \Delta_{a1}, ~
\delta_z \equiv \delta^z_a - \Delta_{a2} = \delta^z_b-\Delta_{b2}
\label{DEL}
\end{equation}
will be used in the following derivations. We can show that the resulting
effective Hamiltonian can be obtained via adiabatic eliminations without
involving the undesired oscillating terms. In fact, under the new conditions
in this case, all of these parameters $\delta_k$ should be much smaller than
the detunings for the laser fields and the cavity modes.

Through two adiabatic elimination processes that are same as in the above
subsections, we can obtain, within the zero-photon subspace, a similar
effective Hamiltonian as in Eq.~(\ref{HE2}), but with different parameters
\begin{eqnarray}
B_s\!&\!=\!&\!\frac{1}{8}\epsilon^s\left[\frac{(\Omega_{b2}\kappa_a^s)^2}{\delta_a^s}-\frac{(\Omega_{a1}\kappa_b^s)^2}{\delta_{b}^s}\right]\,,\nonumber\\ B_z\!&\!=\!&\!\frac{1}{8}\epsilon^z\left[\frac{(\Omega_{b2}\kappa_b^z)^2}{\delta_b}-\frac{(\Omega_{a2}\kappa_a^z)^2}{\delta_{a}}\right]\,,\nonumber\\
J_{s1}\!&\!=\!&\!\frac{t_s}{4}\left[\left(\frac{\epsilon^s\kappa_b^s\Omega_{a1}}{g_b^s}\right)^2+\left(\frac{\epsilon^s\kappa_a^s\Omega_{b2}}{g_a^s}\right)^2\right]\,,\nonumber\\ J_{s2}\!&\!=\!&\! \frac{t_s}{2}\,\frac{(\epsilon^s)^2\kappa_a^s\kappa_b^s\Omega_{a1}\Omega_{b2}}{g_a^sg_b^s}\,,\nonumber\\
J_{z1} \!&\!=\!&\! \frac{t_z}{2}\left(\frac{\epsilon^z\kappa_a^z\Omega_{a2}}{g_a^z}\right)^2\,,~~ J_{z2} =  \frac{t_z}{2}\left(\frac{\epsilon^z\kappa_b^z\Omega_{b2}}{g_b^z}\right)^2\,,\nonumber\\
J_{z3}\!&\!=\!&\!\frac{t_z}{2}\,\frac{(\epsilon^z)^2\kappa_a^z\kappa_b^z \Omega_{a2}\Omega_{b2}}{g_a^zg_b^z}\,\nonumber
\end{eqnarray}
where we have defined $\epsilon^k\equiv1/(1+\delta_k/\lambda_k)$,
$\kappa_a^k\equiv 1+\delta_k/2\delta_a^k$, and
$\kappa_b^k\equiv1+\delta_k/2\delta_b^k$. Now the condition (9) becomes
\begin{equation}
\lambda_k+\delta_k\gg \frac{|g_{\alpha}\Omega_{\beta}|}{4}\left(\frac{1}{\delta_{\alpha}} + \frac{1}{\Delta_{\beta}}\right),~t_k\,.
\label{22aec}
\end{equation}
\begin{center}
\begin{figure}
\includegraphics[width=3.2in]{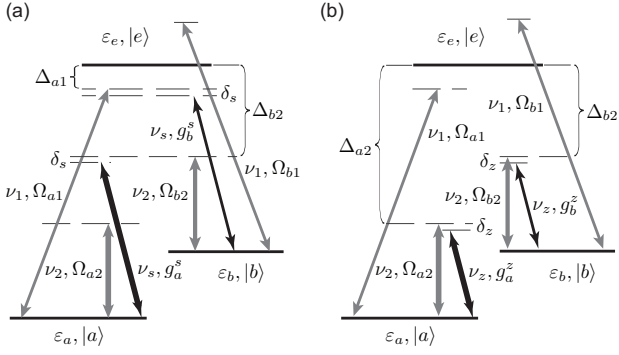}
\caption{(a) Schematic diagram of a $\Lambda$-type three-level atom coupled with two
laser fields and a mode of the cavity in the $s$-bond direction of the
honeycomb lattice, where $s=x$ and $y$. (b) Schematic diagram of a
$\Lambda$-type three-level atom coupled with two laser fields and a mode of
the cavity in the $z$-bond direction of the honeycomb lattice. In both
(a) and (b), we introduce nonzero detuning differences
$\delta_s\equiv\Delta_{a1}-\delta_b^s=\Delta_{b2}-\delta_a^s$ and
$\delta_z\equiv\Delta_{a2}-\delta_a^z=\Delta_{b2}-\delta_b^z$, in contrast
to the case in Figures~2(b) and 2(c).} 
\label{fig4}
\end{figure}
\end{center}

If the parameters are tuned to satisfy the prescribed conditions listed in
Table.~2, we can also derive an effective Hamiltonian for the Kitaev model,
\begin{equation}
H_{{\rm eff}}=J_x^{\prime}\sum_{x-{\rm link}}\sigma^x_i\sigma^x_j + J_y^{\prime}\sum_{y-{\rm link}}\sigma^y_i\sigma^y_j + J_z^{\prime}\sum_{z-{\rm link}}\sigma^z_i\sigma^z_j\,, \label{HEF}
\end{equation}
where
\begin{eqnarray}
J_x^{\prime}\!&\!=\!&\!\frac{t_x}{2}\left(\frac{\epsilon^x\kappa_a^x\Omega_{b2}}{g_a^x}\right)^2\,,~~
J_y^{\prime}=\frac{t_y}{2}\left(\frac{\epsilon^y\kappa_b^y\Omega_{b2}}{g_a^y}\right)^2\,,\nonumber\\
J_z^{\prime}\!&\!=\!&\!\frac{t_z}{2}\left(\frac{\epsilon^z\kappa_b^z\Omega_{b2}}{g_b^z}\right)^2\,,\nonumber
\end{eqnarray}
are the $x$-, $y$-, and $z$-type coupling strengths between atoms at
nearest-neighbor sites when Eq.~(\ref{22aec}) is satisfied.

Similarly, an effective Hamiltonian containing the magnetic term can be obtained
in this more general case by changing the Rabi frequency $\Omega_{a2}$ in the last
column of Table~2,
\begin{eqnarray}
H_{{\rm eff}} \!&\!=\!&\! B^{\prime}\sum_j\sigma^z_j
+ J_x^{\prime}\sum_{x-{\rm link}}\sigma_i^x\sigma_j^x\nonumber\\
\!&\!\!&\!+J_y^{\prime}\sum_{y-{\rm link}}\sigma_i^y\sigma_j^y+J_{zc}^{\prime}\sum_{z-{\rm link}}\sigma_i^z\sigma_j^z\,,
\label{HEFB}
\end{eqnarray}
where $B^{\prime}=B_{z}^{\prime}-(J_{z1}^{\prime}-J_{z2}^{\prime})/4$, and
$J^{\prime}_{zc}=(J_{z1}^{\prime} + J_{z2}^{\prime} - 2J_{z3}^{\prime})/4$.

\begin{center}
\begin{tabular}{p{17.5mm}p{25mm}p{25mm}}
\multicolumn{3}{p{80mm}}{{\bf Table~2} \quad Conditions of the
parameters for implementing the Kitaev model in a more general
case when Eq.~(\ref{22aec}) is satisfied.}\\
\hline\hline
Parameters & \quad$x$-type bond $~~~$ $~~~~~~~$($\sigma^x\sigma^x$) & \quad$y$-type bond $~~~$ $~~~~~~~$($\sigma^y\sigma^y$) \\
\hline
\quad$\Omega_{a1}$ & \multicolumn{2}{c}{} \\
\quad$\Omega_{b2}$ & \multicolumn{2}{c}{\raisebox{1.8ex}[0pt]{$\gamma_s(\kappa_b^s\Omega_{a1})^2=(\kappa_a^s\Omega_{b2})^2$}\quad\quad\quad\quad} \\
\quad$\Omega_{a2}$ & \quad\quad\quad------ & \quad\quad\quad------ \\
\quad$g^k_a$ &\quad\quad $g_a^xg_b^x>0$ &\quad\quad$g_a^yg_b^y<0$ \\
\quad$g^k_b$ & \multicolumn{2}{c}{$(g^k_a)^2=\gamma_k(g^k_b)^2$\quad\quad\quad\quad} \\
\hline\hline
\\
\hline\hline
Parameters & \multicolumn{2}{c}{$z$-type ($\sigma^z\sigma^z$)\quad\quad\quad\quad} \\
\hline
\quad$\Omega_{a1}$ & \multicolumn{2}{c}{------\quad\quad\quad\quad} \\
\quad$\Omega_{b2}$ & \multicolumn{2}{c}{} \\
\quad$\Omega_{a2}$ & \multicolumn{2}{c}{\raisebox{1.8ex}[0pt]{$(\kappa_a^z\Omega_{a2})^2=\gamma_z(\kappa_b^z\Omega_{b2})^2$}\quad\quad\quad\quad} \\
\quad$g^k_a$ & \multicolumn{2}{c}{$g_a^zg_b^z<0$\quad\quad\quad\quad} \\
\quad$g^k_b$ & \multicolumn{2}{c}{$(g^k_a)^2=\gamma_k(g^k_b)^2$\quad\quad\quad\quad} \\
\hline\hline
\end{tabular}
\end{center}

So far, we have shown that the effective Hamiltonian of the cavity-lattice system
with embedded atoms can be reduced to the Kitaev model if we choose proper
detunings and Rabi frequencies for both laser fields and cavity modes.
Note that in the cavity systems, every two adjoining cavities can be
linked by an optical fiber, and the photon-tunnelling rate $t$ between
nearest-neighbor cavities can be continuously tuned by twisting the fibers.

\section{Possible Experimental Implementation}

In this section, we show how the scheme considered above can be achieved using
the currently existing technologies. Central to the success of our scheme is
that the coefficients of the effective Hamiltonians must be much larger than
the decay rates of the cavities and the excited states $|e_j\rangle$. Below we show 
that these requirements are experimentally accessible.

We use the following notations: $\Omega=\max\{\Omega_{a1}$, $\Omega_{a2}$,
$\Omega_{b1}$, $\Omega_{b2}\}$, $g=\max\{g_a^k$, $g_b^k\}$,
$\Delta=\min\{|\Delta_{a1}|$, $|\Delta_{a2}|$, $|\Delta_{b1}|$, $|\Delta_{b2}|\}$,
$\delta=\min\{\mathit{\delta_k}\}$, and $t=\min\{t_k\}$.  The occupation of the
excited states $|e_j\rangle$ can be estimated as~[19]
\begin{equation}
\langle|e_j\rangle\langle e_j|\rangle\approx\left(\frac{\Omega}{\Delta}\right)^2\,,\label{PES}
\end{equation}
and the photon number $n_{{\rm ph}}$ is
\begin{equation}
n_{{\rm ph}}\approx\left(\frac{g\Omega}{\delta\Delta+g^2+\Omega^2/4}\right)^2 \,. \label{PPH}
\end{equation}
The coupling constants between the pseudo-spins are approximately given by
\begin{equation}
J_k,J_k^{\prime}\approx t\left(\frac{g\Omega}{\delta\Delta+g^2}\right)^2\,. \label{CR}
\end{equation}
The two types of the effective decay rates are represented by
$\Gamma_1=(\Omega/\Delta)^2\gamma$, and
$\Gamma_2=[g\Omega/(\delta\Delta+g^2+\Omega^2/4)]^2\kappa$, where $\gamma$ and
$\kappa$ are the rate of the spontaneous emission from the excited level and that
of the cavity decay, respectively.  Hence, the condition
$\Gamma_1, \Gamma_2 \ll [g\Omega/(\delta\Delta + g^2)]^2 t$ is experimentally
realizable when
\begin{equation}
\gamma \ll \mbox{min} \left\{ t\left( \frac{g}{\delta} \right)^2, t\left( \frac{\Delta}{g} \right)^2 \right\}, \quad \mbox{and} \quad \kappa \ll t\,. \label{CON1}
\end{equation}

Furthermore, the assumption (13) can also be satisfied in our scheme. By defining
$\mu\equiv |g/\Omega|$ and $\eta\equiv|\delta\Delta/g\Omega|$, this inequality
approximately reduces to $\mu+\eta\gg1/2$, which can be satisfied if
$2|\delta \Delta| \gg |g\Omega|$ or $2|g| \gg| \Omega|$. In the ordinary case,
detunings can be adjusted to be larger than gigahertz, while $g$ and $\Omega$ are
around hundreds of megahertz for strongly coupled cavity-atom systems. Then the
first inequality, i.e., $2|\delta \Delta| \gg |g\Omega|$, can thus be satisfied.
On the other hand, for ultrastongly coupled cavity-atom systems, such as photonic
band gap cavities~\cite{KH}, where $g$ can reach $\sim 20$ GHz, the second inequality,
i.e., $2|g| \gg| \Omega|$, can also be fulfilled.

The above arguments suggest that cavities with a high cooperativity factor and a
high cavity-atom coupling strength $g$ are desirable for implementing the Kitaev
model. Here the cooperativity describes the loss from the atom and the cavity, and
is defined as $C=g^2/\kappa\gamma$. Encouragingly, cavities with the required
properties can be realized in certain high-$g$ microcavities with current
micro-manufacturing and micro-etching technologies (see reference~\cite{SMS} and
references therein).  For example, in toroidal microcavities, the cooperativity
factor of $C \sim 10^7$ was achieved experimentally~\cite{SMS}, and in the photonic band
gap cavities,  $C \sim 10^3$ was also realized~\cite{BSS,KH,JV}. In order to achieve the
honeycomb lattice, one can build three optical cavities in different directions at
each site of the lattice, or design an appropriate structure of the photonic band
gap cavities to restrict the photons of different frequencies in three directions.
However, some microcavities have only one mode at each site of the cavity lattice,
such as the toroidal cavities~\cite{SMS}, which are not good candidates for the
implementation of the Kitaev model. Alternatively, such cavity systems may be
useful for simulating other spin models, e.g., Heisenberg spin models~\cite{MJH1,JC1,CZX}. In
summary, we highlight that the high-finesse optical cavities (such as Fabry-Perot
cavities)~\cite{CJH} and photonic band gap cavities~\cite{ADG,BSS} are two promising candidates
to implement our scheme in 2D lattice of microcavities, owing to the possibility of
designing an appropriate geometrical structure for different cavity modes.

\section{Discussion and conclusion}

Here we emphasize that the different couplings of the effective Hamiltonians
can be simultaneously achieved by using suitable cavity modes depending on
the geometrical structure of the considered cavity lattice and by properly
applying detuned laser fields. Our scheme can also provide a model system
for observing the phase transition between the gapless phase and gapped
phase of the Kitaev model by tuning the external laser fields. In our proposal,
for simplicity, all the parameters for the laser fields are assumed to be
identical. However, if one can control the laser fields at each site of the
honeycomb lattice, the model described in this paper can be realized with
more flxibilities, owing to the tunable parameters of the system.

In conclusion, we have proposed an approach to realize the Kitaev model by
using an anisotropic cavity lattice with embedded field-controlled
$\Lambda$-type three-level atoms. Our approach provides an artificial, but
experimentally realizable, many-body spin system for demonstrating the
topological phases in the Kitaev model. Also, our approach can be extended
to simulate other 2D and even 3D spin models.

\acknowledgements

This work was supported by the National Basic Research Program of China Grant No. 2009CB929302, the National Natural Science Foundation of China Grant No.91121015, the MOE of China Grant No. B06011, and the NSF PHY-0925174.

%\textit{et al.}

\end{document}